\documentclass[twocolumn,apl,amsmath,amssymb,showpacs,superscriptaddress]{revtex4-2}
\usepackage{epsf}      
\usepackage{graphicx}
\usepackage{color}
\usepackage{soul}
\usepackage{gensymb}
\usepackage{sidecap}
\usepackage{amsmath}
\usepackage{mathtools}
\usepackage{float}
\usepackage[hidelinks,colorlinks=true,linkcolor=blue,citecolor=blue]{hyperref}

\begin{document}
\title{Dielectric behavior and impedance spectroscopy of Niobium substituted Lanthanum based orthovanadates at high temperatures}

\author{Ashok Kumar}
\affiliation{Department of Applied Physics, Delhi Technological University, Delhi-110042, India}
\affiliation{Department of Physics, Atma Ram Sanatan Dharma College, University of Delhi, New Delhi-110021, India}
\author{Vikas N. Thakur}
\email{Present Address: Department of Physics, School of Physical Sciences, DIT University, Dehradun 248009, India}
\affiliation{Department of Physics, Indian Institute of Technology Delhi, Hauz Khas, New Delhi-110016, India}
\author{Ajay Kumar}
\email{Present Address: Ames National Laboratory, U.S. Department of Energy, Iowa State University, Ames, Iowa 50011, USA}
\affiliation{Department of Physics, Indian Institute of Technology Delhi, Hauz Khas, New Delhi-110016, India}
\author{Vinod Singh}
\affiliation{Department of Applied Physics, Delhi Technological University, Delhi-110042, India}
\author{Anita Dhaka}
\affiliation{Department of Physics, Hindu College, University of Delhi, New Delhi-110007, India}
\author{Rajendra S. Dhaka}
\email{Corresponding author: rsdhaka@physics.iitd.ac.in}
\affiliation{Department of Physics, Indian Institute of Technology Delhi, Hauz Khas, New Delhi-110016, India}

\date{\today}      

\begin{abstract}

We present the detailed study of the temperature dependent dielectric properties, impedance spectroscopy and electrical conductivity of LaV$_{1-x}$Nb$_x$O$_4$ ($x=$ 0--1) samples prepared by the solid-state reaction method. The dielectric constant ($\epsilon_r'$) increases (decreases) with increase in the temperature (frequency); while, the magnitude of $\epsilon_r'$ remains almost invariant (order of 10$^4$ at 100~Hz and 600\degree C) with $x$. The single phase $x=$ 0 (monoclinic-monazite, $P2_1/n$) and $x =$ 1 (monoclinic fergusonite, $I2/a$) samples show the lower values of loss factor [tan$\delta=$ (4--8)] as compared to the $x=$ 0--0.8 samples [tan$\delta=$(12--18)] having mixed tetragonal and monoclinic phases, which indicates the strong correlation between the crystal structure and the dielectric properties. The real part of impedance ($Z'$) decreases with both temperature and frequency, and the observed weak relaxation remains almost unaltered with $x$. The imaginary part of impedance ($Z''$) shows strong relaxation peaks shifting towards higher temperatures with frequency, which is attributed to the effect of grains, grain boundaries, and electrodes in the samples. The activation energy of the relaxation process is estimated to be 0.8--1.0~eV for the $x=$ 0--0.6 samples, and $\approx$1.4~eV for the $x=$ 0.8; whereas the $x=$ 1 sample shows two values ($\approx$0.5~eV and $\approx$1.0~eV) in the higher and lower temperature range, respectively. Further, the change in total conductivity with the angular frequency, which found to be in the range of 10$^{-3}$ to 10$^{-5}$~S/m, is fitted using the Jonsher power law. The analysis suggests the overlapping large polaron tunneling (OLPT) model for all the samples, whereas the $x$ = 1 sample exhibit a transition near 480\degree C and at higher temperatures it shows non-overlapping small polaron tunneling (NSPT) and quantum mechanical tunneling (QMT). This transition is corroborated by the tangent loss curves and may be associated with the change in structure.  
\end{abstract}

\maketitle
\section{\noindent ~Introduction}

In last few decades, the orthovanadates with general formula RVO$_4$, where R refers to rare earth elements, have widely attracted the attention of scientific community due to their vast technological applications in catalysts, as luminescent materials, polarizers, and laser host materials \cite{XiaJAP00}. In this family, the LaVO$_4$ is turned out to be particularly important in the recent time because of its use as an electrode material for Lithium-ion batteries (LIBs) as it shows large specific capacity as well as long cyclic life \cite{YiJAL17}.  The LaVO$_4$ crystallises in tetragonal zircon type polymorphs with space group I4$_1$/amd and monoclinic monazite type polymorphs with space group P2$_1$/n, following the trend of the rare-earth family \cite{Ashok_Jalcom_23}. However, it also deviates from the trend by thermally stabilizing in the monazite type structure and being metastable at ambient temperatures \cite{Ashok_Jalcom_23}. This is due to the higher oxygen coordination number in the monazite as compared to zircon type structure and largest ionic radius of La$^{3+}$ in the lanthanide family \cite{InguscioPP93, AxelS00, SunJAP10}. A pattern of VO$_{4}$ tetrahedra with four distinct V--O bonds and RO$_8$ dodecahedra sharing edges alternately and connected in chains along the $c$-axis can be found in the zircon structure \cite{RiceACB76}. In the monazite structure, RO$_9$ polyhedra are linked to deformed VO$_{4}$ tetrahedra with four distinct V--O links, sharing their edges \cite{ChakoumkosJSSC94}. According to the refs.~\cite{ZabranskyJPCR05} and \cite{LiuJAC12}, the Zircon type LaVO$_{4}$ cannot be synthesized at ambient temperatures using the traditional solid state reaction technique; however, it can be produced and stabilized using the hydrothermal method.

The influence of substituting La by another rare earth elements has been significantly explored for the LaVO$_{4}$ family; however, the doping/substation at the V-site is still in its infancy, as evidenced by the fact that only a handful of studies have been conducted so far \cite{VermaACAG01, HimanshuPRB21, Ashok_Jalcom_23}. The rare-earth niobate LaNbO$_{4}$ on the other hand is well-known to show the temperature and chemical pressure (doping) induced structural transformation \cite{TakeiJCG77, AldredML83}. The LaNbO$_4$ transforms from the monoclinic fergusonite (space group $I2/a$) structure at the low temperature to the tetragonal scheelite (space group $I4_1/a$) phase above $\sim$495$\degree$C \cite{TakeiJCG77}. Similarly, it also shows a structural transformation on replacing 25\% of Nb$^{5+}$ with V$^{5+}$ \cite{AldredML83}. Because of its intriguing characteristics, including proton conductivity \cite{HaugsrudNM06}, excellent dielectric, and high energy release using X--ray stimulation \cite{BlasseCPL90}and hence wide technological applications in sensors \cite{BalamuruganASS13}, contrast agents, waveguides, ferroelectrics \cite{GuFE83}, phosphors \cite{LiPCCP15}, laser crystals \cite{DingRSCA17}, scintillators \cite{PorterNSSCR06}, luminophores, LEDs, etc., the LaNbO$_{4}$ is getting a special attention of the scientific community from several decades. Therefore, we choose to replace V by Nb atoms in LaV$_{1-x}$Nb$_x$O$_4$, as Nb shares many properties with vanadium and is found right beneath it in the periodic chart, to examine the influence of B-site atoms on the high temperature structural transition and hence the physical properties of these compounds. Though, the vanadium costs have historically lagged behind those of niobium, but it has lately increased by about 300\%, which has created strong economic incentives to transition to niobium. The tetrahedrally coordinated niobium has bigger ionic radius (0.48 \AA), and is isoelectronic to the vanadium ion (V$^{+5}$) \cite{ErrandoneaPMS08} and hence most suitable to exert the chemical pressure in the crystal lattice. Further, the nontoxic nature of Nb make it suitable candidate for the industrial applications. Moreover, the La-Nb based perovskite oxides have been also extensively studied due to their important fundamental as well as functional properties \cite{AjayPRB20, AjayPRB2, AjayPRB22, 39, ShuklaJPCC19, ShuklaJPCC21, ThakurMRB01}. 

Recently, we report the vibrational, structural as well as electronic properties of the LaV$_{1-x}$Nb$_x$O$_4$ ($x =$0--1) samples using the x-ray diffraction (XRD), Raman and x-ray photoemission spectroscopy \cite{Ashok_Jalcom_23}.  The $x=$ 0 and 1 samples were found to crystallize in the single phase monoclinic-monazite ($P2_1/n$) and monoclinic fergusonite ($I2/a$) type structures at the room temperature, respectively; whereas the monoclinic-monazite ($P2_1/n$) and tetragonal scheelite ($I4_1/a$) phases coexist in the $x=$ 0.2--0.8 samples and phase fraction of later increases with the Nb substitution \cite{Ashok_Jalcom_23}. The fact that the  LaNbO$_4$ possesses the tetragonal structure when measured at room temperature by substituting about 20$\%$ of Nb$^{5+}$ concentration at the V$^{5+}$ site is an important finding, which makes a variety of room-temperature applications possible, as this transformation plays a significant role in regulating the protonic conductivity of LaNbO$_4$ \cite{HuseJSSC12}. Moreover, for certain compositions of LaV$_{1-x}$Nb$_x$O$_4$, the transition temperature can occur close to room temperature \cite{AldredML83}. At $x$=0.75, for instance, temperature-dependent XRD measurements indicate that LaV$_{1-x}$Nb$_x$O$_4$ should possess a tetragonal structure when measured at room temperature, with its transition at around 250~K. Nevertheless, the XRD pattern below this transition also shows minor residual intensity in the form of broadened lines, which may be the effect of precursors \cite{AldredML83}. Similarly, the XRD patterns of $x$=0.8 samples also exhibit broad peaks, likely due to the same effect \cite{AldredML83}. Therefore, it is vital to investigate the high temperature phase transition and their influence on the dielectric and ac impedance spectroscopy with Nb substitution in LaV$_{1-x}$Nb$_x$O$_4$ samples. The impedance spectroscopy is widely used to study the electrical properties of materials to understand their various possible applications such as solid oxide fuel cells (SOFCs). One of the advantages of impedance spectroscopy is its ability to provide information about the properties of a material or system over a wide frequency range, typically from a few Hertz to several mega Hertz. This can be particularly useful for studying materials that exhibit complex electrical behavior, such as those with multiple structural phases or with a distribution of relaxation times \cite{NagaoPRB07, 54, 57, 52}.  

Therefore, in this article, our objectives are to comprehensively investigate the temperature dependent conductivity, impedance spectroscopy, and dielectric properties of LaV$_{1-x}$Nb$_x$O$_4$ ($x =$0--1) samples. These measurements are made of the dielectric constant and tangent loss at temperatures ranging from 25\degree C to 600\degree C and in a wide range of frequency ($f$) from 100 Hz to 1 MHz. We demonstrate that the dielectric constant increases with temperature and decreases in magnitude with frequency. The magnitude of the dielectric constant is not significantly changed by the addition of Nb$^{5+}$ in place of V$^{5+}$, and anomalous behaviour is observed from $x$ = 0.2 to 0.8 due to their mixed states, whereas the $x$ = 0 sample shows no peak and the $x$ = 1 sample shows a weak relaxation in higher temperature range. For the $x$ = 0--0.8 samples, the tangent loss displays the peaks in the higher temperature range above 360\degree C, whereas the $x$ = 1 displays a weak relaxation peak, shifting towards higher temperatures. The real component of the impedance ($Z'$) decreases with temperature and frequency, and a weak relaxation is observed in all the samples, which remains almost invariant with $x$. The imaginary component of impedance ($Z''$) exhibits prominent relaxation peaks that shift towards higher temperatures with frequency, which further illuminating the influence of the grain, grain boundary, and electrode in the samples. In the $Z''$--$f$ plots, the behaviour from $x$ = 0 to 0.6 is identical, but the $x$ = 0.8 and 1 samples exhibit a relaxation in the lower frequency area that exhibits both the grain effect and the grain boundary effect. In order to look at the lower frequency components, such as the grain boundary and electrode influence, the modulous spectroscopy is also used. Additionally, the total conductivity is measured with varied frequency in the 360\degree C--560\degree C temperature range and fitted with the Jonsher power law. The $x$ = 0.8 sample shows the lowest conductivity, which is related to the largest grain size and highest Nb concentration in the coexisting phases of these samples. Interestingly, we find the overlapping large polaron tunneling (OLPT) behavior for all the samples, whereas the $x$ = 1 sample undergoes non-overlapping small polaron tunneling (NSPT) and quantum mechanical tunneling (QMT) behavior at higher temperatures. 

\section{\noindent ~Experimental}

We use the conventional solid state reaction method to prepare the LaV$_{1-x}$Nb$_x$O$_4$ ($x$ = 0--1, $\Delta$x=0.2) bulk samples. The details of the synthesis process, and their structural and electronic properties can be found in our earlier report \cite{Ashok_Jalcom_23}. The prepared samples were cold pressed in the form of circular pellets of 10 mm diameter and 1-2 mm thickness at 14~MPa before the final annealing at 1250\degree C. The pellets were then polished and painted with sliver paste on both the sides for the dielectric measurements. The temperature dependent dielectric measurements and ac impedance spectroscopy were then performed on all the samples using HIOKI 3536 LCR meter from 25\degree C to 600\degree C in the large frequency range from 100 Hz to 1 MHz at 1 V excitation voltage, under the vacuum conditions of $\sim$3$\times$10$^{-2}$~mbar. 

The complex dielectric constant ($\epsilon_r$), tangent loss (tan$\delta$), complex electric modulus ($M$), and ac conductivity are calculated using the temperature dependent complex impedance ($Z$) and phase angle ($\theta$) measured using the LCR meter. It involves applying a small, alternating voltage to a sample and measuring the resulting current response over a range of frequencies. By analyzing the relationship between the current and voltage, the impedance of the sample can be determined. We use the following formulas to calculate various electrical parameters:

\begin{equation}
Z = Z'+iZ''
\end{equation} 

\begin{equation}
M =i\omega C_0Z= M'+iM'',
\end{equation} 

where, $Z'$=$\lvert Z \rvert cos\theta$ and $Z''$=$\lvert Z \rvert sin\theta$ are real and imaginary parts of complex impedance, $M'$ and $M''$ are real and imaginary parts of complex electric modulus, $\omega$=2$\pi f$ is the angular frequency, $C_0$ is the capacitance in the vacuum, defined as $C_0$=$A$$\epsilon_0$/$d$, where $A$ is the area of electrodes and $d$ is the separation between them.

Using equations (1) and (2), the $M'$ and $M''$ can be expressed as: 

\begin{equation}
M'=\omega C_0Z'';  M''=\omega C_0Z'
\end{equation} 

Here, the dielectric constant can be written in terms of the inverse of electric modulus, i.e.,

 \begin{equation}
\epsilon_r=\epsilon_r'+i\epsilon_r''= 1/M
\end{equation}

where the $\epsilon_r'$ and $\epsilon_r''$ are real and imaginary parts of $\epsilon_r$, respectively.

Using equations (2), (3) and (4), we get 

 \begin{equation}
\epsilon_r'=Z''/\omega C_0Z^2 ; \epsilon_r''=Z'/\omega C_0Z^2
\end{equation}

We utilize the above equations to compute the $\epsilon_r'$, $Z'$, $Z''$, and $M''$ parameters, as discussed in detail below. 

\section{\noindent ~Results and discussion}

 \begin{figure} 
\includegraphics[width=3.5in]{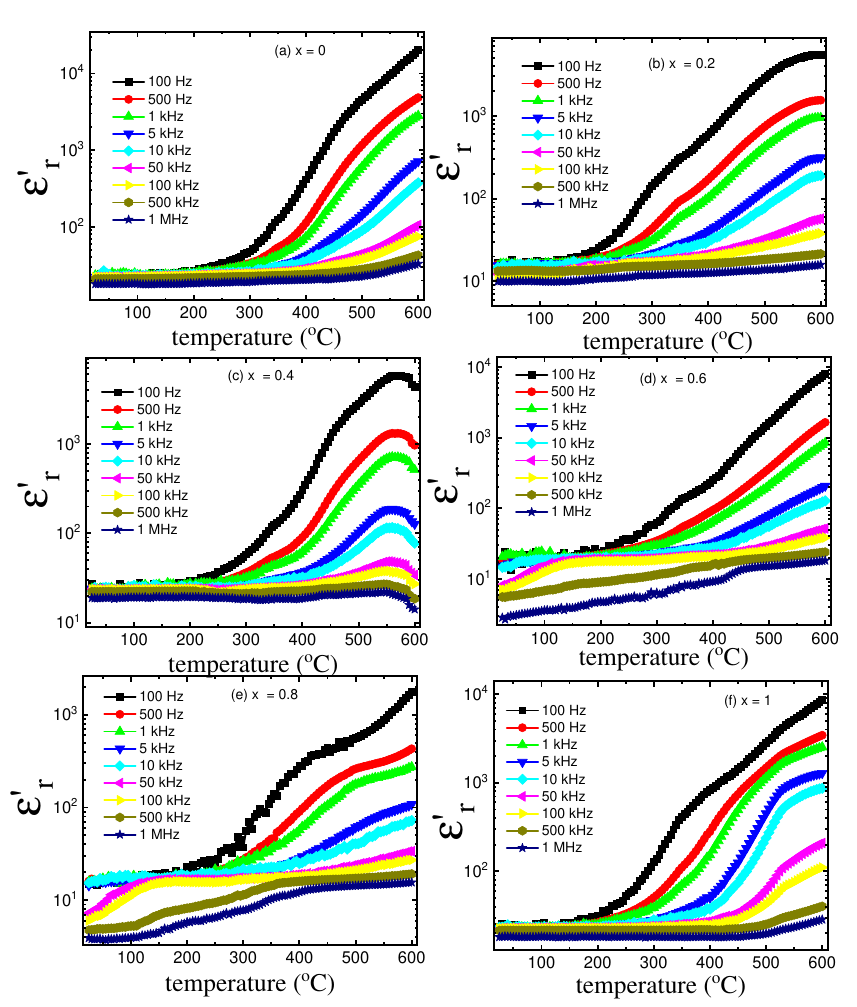}
\caption{(a--f) The temperature dependent real part of dielectric constant ($\epsilon_r'$) for LaV$_{1-x}$Nb$_x$O$_4$ ($x$ = 0--1) samples from 25\degree C--600\degree C in the frequency range of 100 Hz--1 MHz.} 
\label{fit_1}
\end{figure}

\begin{figure*} 
\includegraphics[width=6.4in]{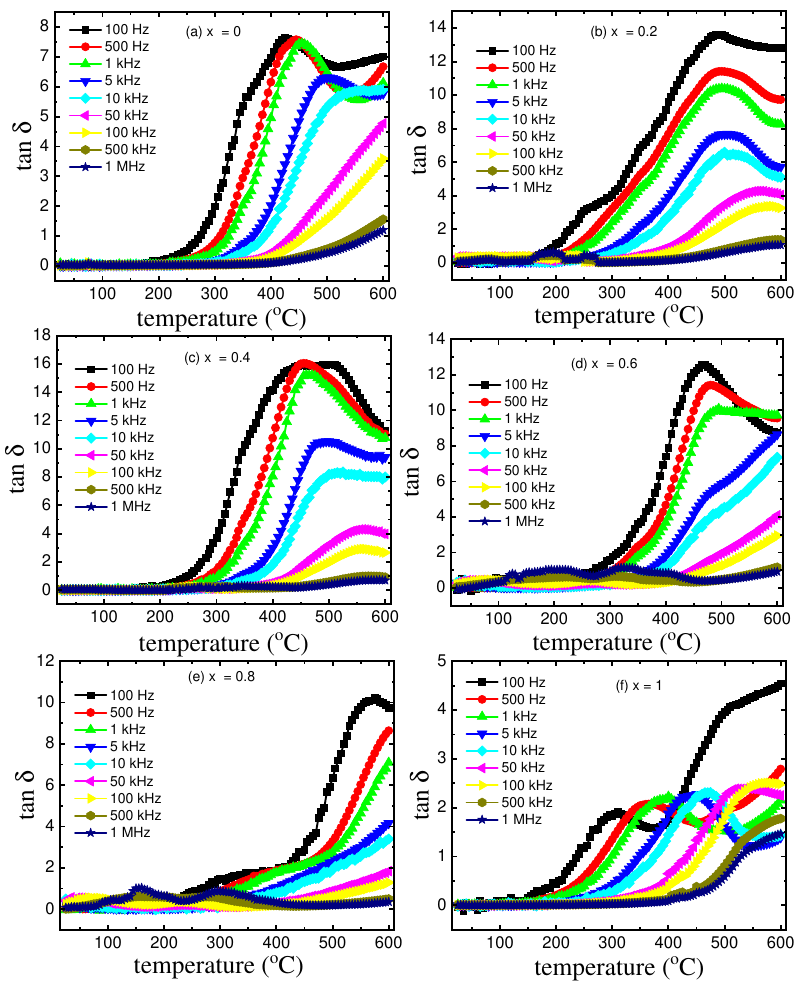}
\caption{(a--f) The dielectric loss (tan $\delta$) for LaV$_{1-x}$Nb$_x$O$_4$ ($x$ = 0--1) samples plotted as a function of temperature at the selected frequencies.} 
\label{fit_2}
\end{figure*}

The temperature dependent real part of dielectric constant ($\epsilon_r'$) is presented in Figs.~\ref{fit_1}(a--f) for LaV$_{1-x}$Nb$_x$O$_4$ ($x =$0--1) samples at the selected frequencies. The dielectric constant of a material determines the amount of electric charge that can be stored in a capacitor made from the material, as well as the strength of the electric field that can be sustained before the material breaks down. For the $x=$ 0 sample, no peak is observed in the measured temperature range and the $\epsilon_r'$ decreases with increase in the frequency [see Fig.~\ref{fit_1}(a)]. The $\epsilon_r'$ remains almost temperature independent up to around 300\degree C and then increases abruptly with further increase in the temperature. The decrease in the $\epsilon_r'$ with frequency is because of the reduction of space charge polarization \cite{53, GaoCPB19, GaoCI18, GaoJALCOM19}. The presence of space charge polarization at lower frequency creates a potential barrier which leads to the building-up of charges at the grain boundary, resulting the higher values of $\epsilon_r'$ \cite{53}. It is conventional to use the thermal energy to demonstrate the relation between the dielectric constant and the change in the temperature \cite{26}. The mobility of charge carriers and, therefore, their rate of hopping are improved at higher temperatures by the additional thermal energy, which is insufficient at the lower temperatures, causing the observed enhancement in the $\epsilon_r'$ with temperature. The dielectric polarization increases with increase in the temperature, which enhances the $\epsilon_r'$ at higher temperatures up to a certain limit, called 'phase transition temperature' and at this temperature the polarization decreases to almost zero, therefore the $\epsilon_r'$ starts decreasing with temperature \cite{Meena_Ceramic_2022, Rayssi_RSC_2018}. However, no phase transition has been observed for the $x =$0 sample in the measured temperature range (up to 600\degree C). Similarly, for the $x =$ 0.2 sample, no peak is observed; however, the curves start saturating near 600\degree C, as shown in Fig.~\ref{fit_1}(b). The substitution of Nb$^{5+}$ may increases the conductivity of the sample, which led to decrease in the phase transition temperature and hence saturation in the dielectric constant. Moreover, a peak is observed near 550\degree C with further increase in the Nb substitution for the $x =$ 0.4 sample [see Fig.~\ref{fit_1}(c)], the position of which remains almost invariant with increase in the frequency. Interestingly, for the $x= $0.6 and 0.8 samples, an additional dielectric relaxation peak is observed at significantly lower temperatures (200--300\degree C), which shifts to the higher values with increase in the excitation frequencies [see Figs.~1(d, e)]. The increase in the tetragonal scheelite (I4$_1/a$) phases with the Nb substitution is expected to give rise to this additional relaxation peak in these samples \cite{Ashok_Jalcom_23}. Moreover, the $x =$ 1 sample shows a similar behavior to the $x =$ 0 [except a weak relaxation around 500\degree C; see Fig.~\ref{fit_1}(f)], which can be attributed to the presence of single phase in both of these samples \cite{Ashok_Jalcom_23}. 

\begin{figure}[h] 
\includegraphics[width=3.4in,height=4.4in]{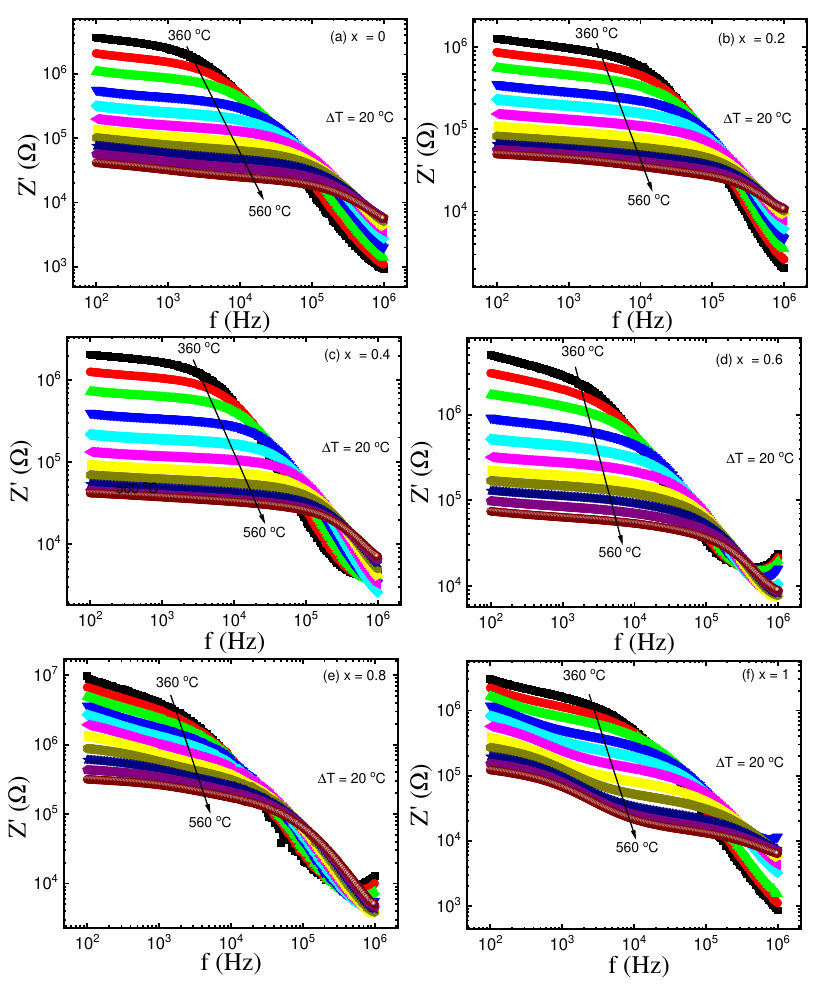}
\caption {(a--f) The real part of impedance ($Z'$) of LaV$_{1-x}$Nb$_x$O$_4$ ($x$ = 0--1) samples plotted as a function of frequency from 100 Hz to 1 MHz at the selected temperature from 360\degree C--560\degree C.} 
\label{fit_3}
\end{figure}

To gain a deeper understanding of the material's behavior, we perform detailed analysis of the temperature-dependent tangent dielectric loss, represented as tan $\delta = \epsilon_r''/\epsilon_r'$, for all the samples, as shown in Figs.~\ref{fit_2}(a--f). Here, we investigate how different samples respond to changes in temperature and excitation frequency by analyzing their tangent dielectric loss, which is a measure of how the material dissipates energy when subjected to an alternating electric field. Note that all the samples in this study exhibit peaks in their tan $\delta$ curves, which are a significant indicator of how the material responds to changes with the variation in temperature and excitation frequency. The location of the peak in tan$\delta$ shifts to higher temperatures as the excitation frequency increases. This shift implies that with the application of higher frequencies, the material's response to temperature changes at higher temperatures. It is an important behavior to note because it indicates a thermally triggered relaxation process \cite{28}. The magnitude of the peak in tan$\delta$ decreases as the excitation frequency increases for most of the samples. This behavior suggests that at higher frequencies, the material's dielectric loss is reduced. However, there is an exception with the sample where $x=$ 1, where the height of the peak in tan$\delta$ actually increases with the frequency until it reaches around 100~kHz. After this point, it starts to decrease as the frequency increases further [see Fig.~\ref{fit_2}(f)]. 

It should be noted that the increase in tan$\delta$ at higher temperatures is primarily due to the direct current ({\it dc}) conductivity. In other words, the behavior of the material at high temperatures is influenced significantly by its ability to conduct the electric current. Interestingly, the tan$\delta$ values increases with Nb substitution up to $x=$ 0.4 sample, and then start decreasing at higher Nb concentration. It is observed that the samples with single phases (with $x=$ 0 and $x=$ 1) have lower dielectric loss values compared to those with mixed phases (with $x=$ 0.2--0.8). This indicates that the presence of mixed phases affects the dielectric properties of the material. For the samples with mixed phases (specifically, $x=$ 0.2--0.8), the interaction between the different phases results in anomalous behavior in the tan$\delta$ curves. This anomalous behavior is observed in higher frequencies and lower temperature regimes \cite{Ashok_Jalcom_23}. It suggests that the combination of different phases introduces complex interactions that affect how the material responds to changes in temperature and excitation frequency. The observed shifts in peak positions, frequency-dependent magnitudes, and differences between single-phase and mixed-phase samples provide insights into the material's electrical properties and its complex behavior under varying conditions. These findings are important for understanding the material's suitability for specific applications and may have implications for the design of electronic devices. From the above analysis and discussion we reveal that the samples with mixed phases ($x$ = 0.2--0.8) exhibit anomalous behavior in their $\epsilon_r'$ and tan$\delta$ characteristics. The presence of mixed phases introduces complex interactions that govern how the electric dipoles respond to the change in temperature and excitation frequency. 

\begin{figure*}
\includegraphics[width=6.7in]{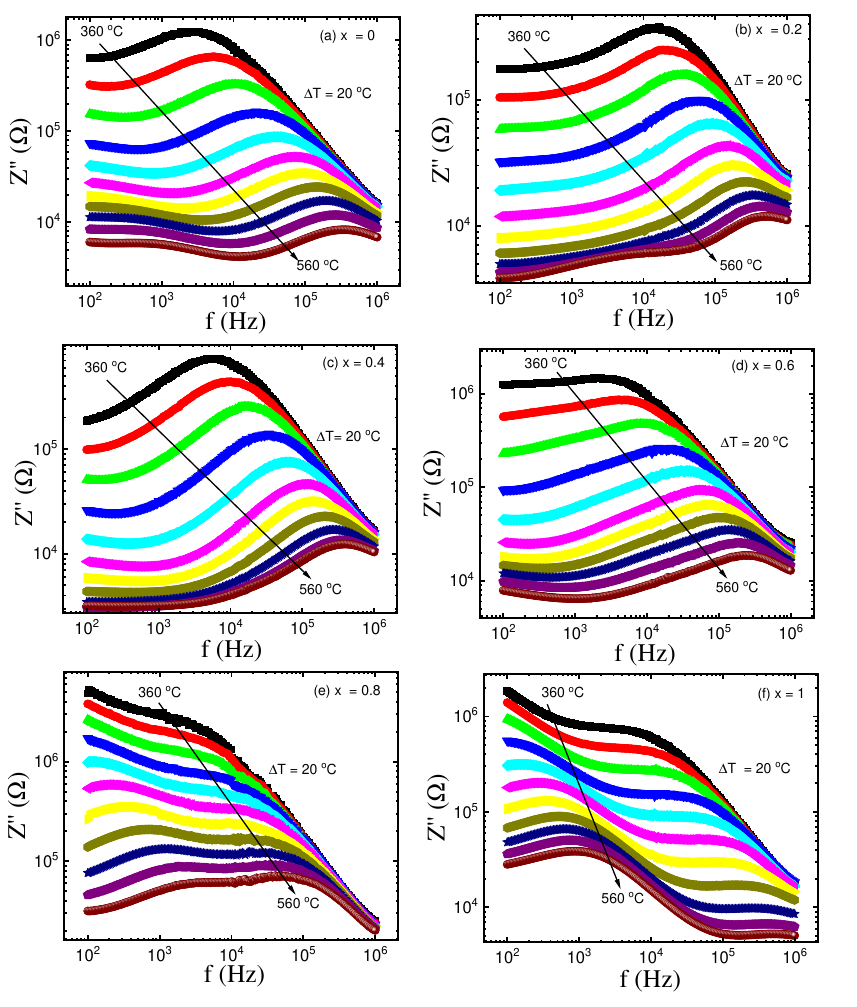}
\caption {(a--f) The imaginary part of impedance ($Z''$) of LaV$_{1-x}$Nb$_x$O$_4$ ($x$ = 0--1) samples plotted as a function of frequency from from 100 Hz to 1 MHz at the selected temperatures from 360\degree C--560\degree C.}
\label{fit_4}
\end{figure*}

\begin{figure*}
\includegraphics[width=6.9in]{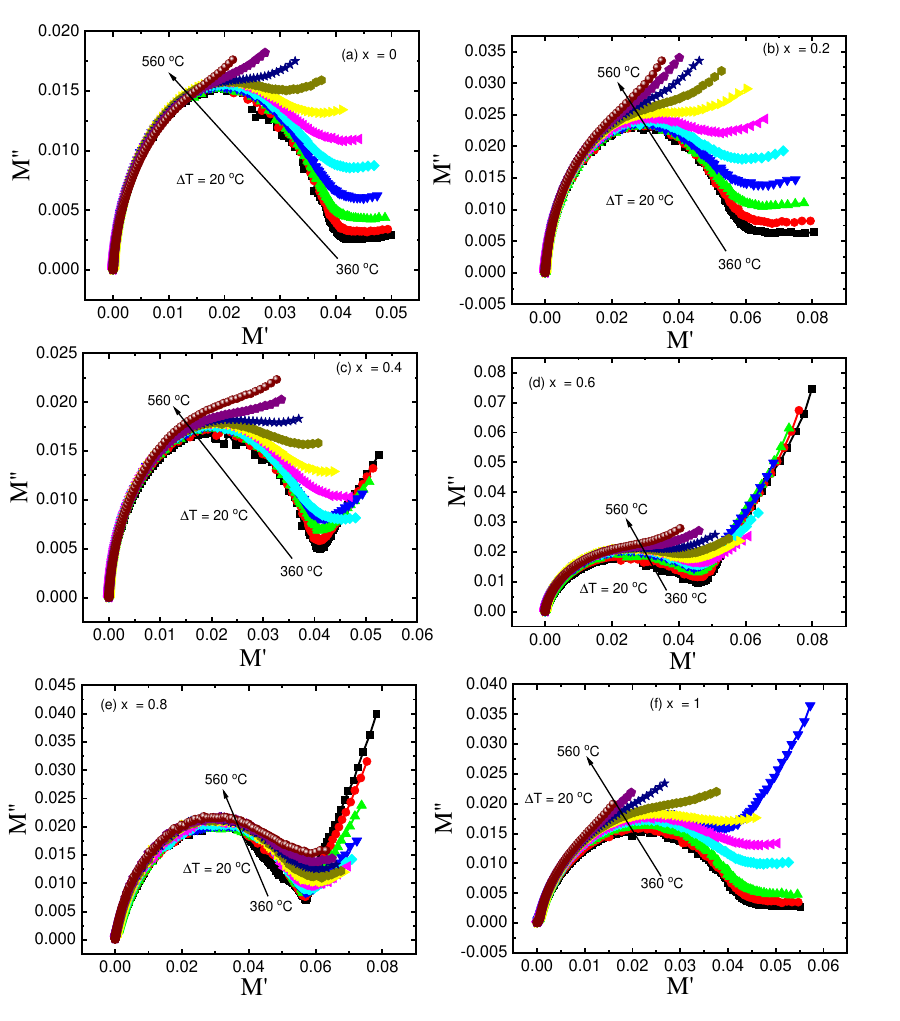}
\caption {(a-f) The imaginary ($M''$) versus real ($M'$) parts of electric modulus as a function of frequency from 100 Hz to 1 MHz at various temperatures from 360 \degree C to 560 \degree C for the LaV$_{1-x}$Nb$_x$O$_4$ ($x$ = 0--1) samples.}
\label{fit_6}
\end{figure*}

Now, we further delved insight into the investigation by analyzing the impedance ($Z$) and modulus ($M$) spectra of all the samples. The aim is to unravel how the grains (individual crystal structures) and grain boundaries (interfaces between these crystal structures) contribute to the observed relaxation processes in the $\epsilon_r'$(T) (real part of dielectric permittivity as a function of temperature) and tan$\delta$(T) (tangent of dielectric loss as a function of temperature) spectra. From the temperature-dependent data of $\epsilon_r'$ and $tan\delta$ in the temperature range of 25 to 600\degree C for all the samples in Figures~1 and 2, respectively, we observed that the significant and rapid increase in the dielectric constant is within the range of 360 \degree C to 560 \degree C. Also, this temperature range exhibited the most pronounced peaks in the tangent loss for all the samples. Therefore, we focus our analysis solely in this specific temperature range (360 \degree C to 560 \degree C) to investigate parameters such as impedance, electric modulus, and conductivity. This approach allow us to gain insights into the behavior of the grain and the effects on grain boundaries resulting from Nb doping in LaV$_{1-x}$Nb$_{x}$O$_4$. These findings provide valuable information about the electrical properties of the material and how they relate to its microstructure, which is crucial for understanding its potential applications. Figs.~\ref{fit_3}(a--f) display the real part of impedance ($Z'$) as a function of frequency for all the samples at selected temperatures. The $Z'$ is a measure of the opposition that a material presents to the flow of alternating current. The key observation is that the order of $Z'$ remains nearly constant regardless of the Nb substitution in the samples. However, the value of $Z'$ decreases with increasing temperature for all the samples, which is indicative of the semiconducting or insulating nature of these materials. In other words, as the temperature rises, the ability of the material to conduct electric current decreases. The impedance behavior can be divided into two regions: (1) frequency-independent region and (2) frequency-dependent region. The first region is associated with {\it dc} conductivity, which means that the charge carriers (electrons or ions) can move over extended distances successfully through a hopping mechanism. In this process, the neighboring charge carriers relax to the location of the moving carrier. This mechanism leads to the {\it dc} conductivity at low frequencies. The second region is linked to the alternating current ({\it ac}) conductivity. Here, the charge carriers have limited or localized movement and fail to hop effectively, instead, they return to their original positions. This mechanism results in dispersive {\it ac} conductivity at higher frequencies. We find that the {\it ac} conductivity dominates above a critical frequency and this critical frequency is determined by the sample's temperature and remains nearly constant across different values of the Nb concentration ($x$). In other words, the transition from {\it dc} to {\it ac} conductivity occurs at a specific frequency that depends on temperature, but found to be independent of $x$ [see Figs.~\ref{fit_3}(a--f)]. \par 

\begin{figure}[h] 
\includegraphics[width=3.5in]{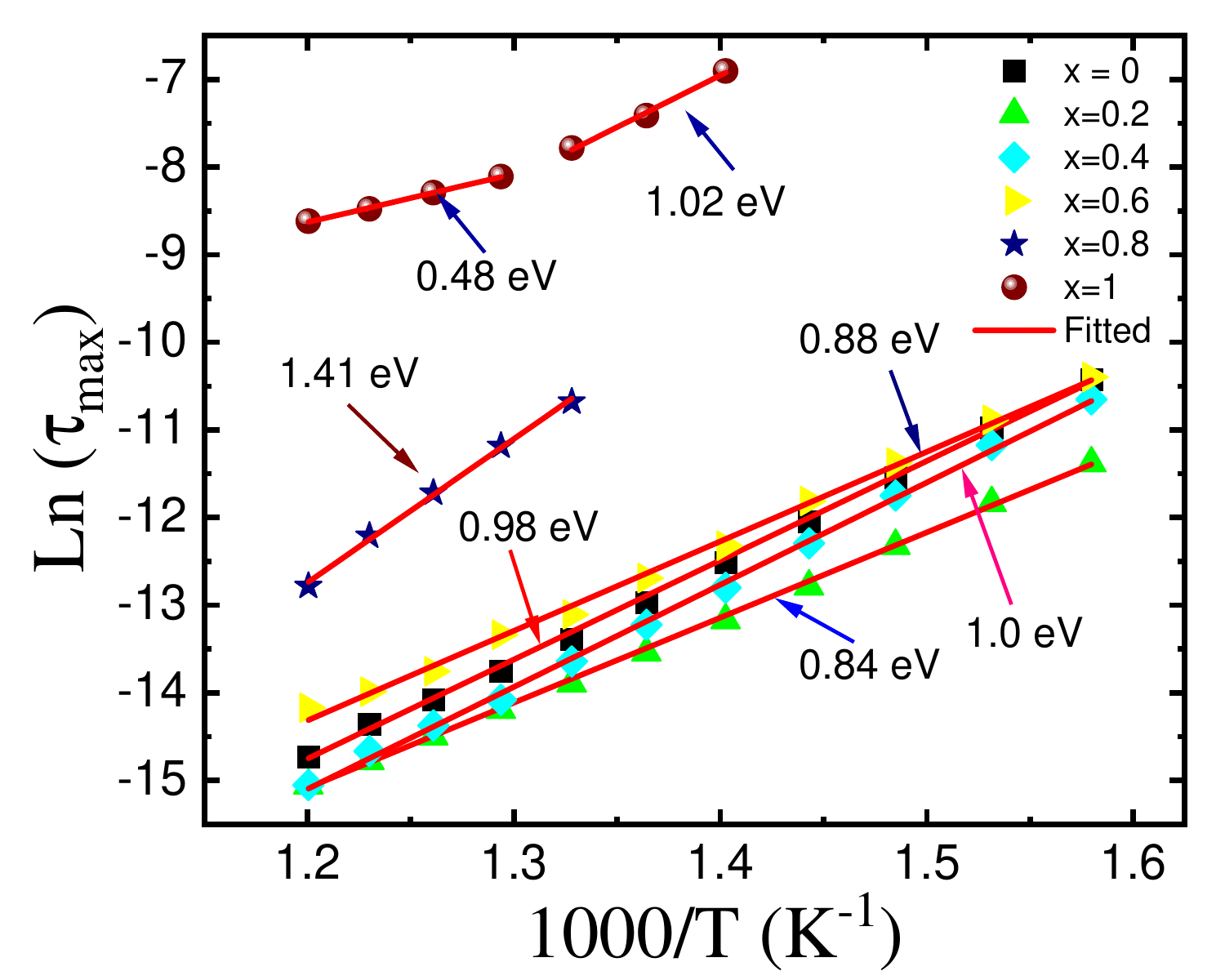}
\caption {The variation in the relaxation time with inverse of temperature for LaV$_{1-x}$Nb$_x$O$_4$ ($x$ = 0--1) samples. The solid red lines represent the best fit using the Arrhenius equation.}
\label{fit_5}
\end{figure}

\begin{figure} 
\includegraphics[width=3.5in]{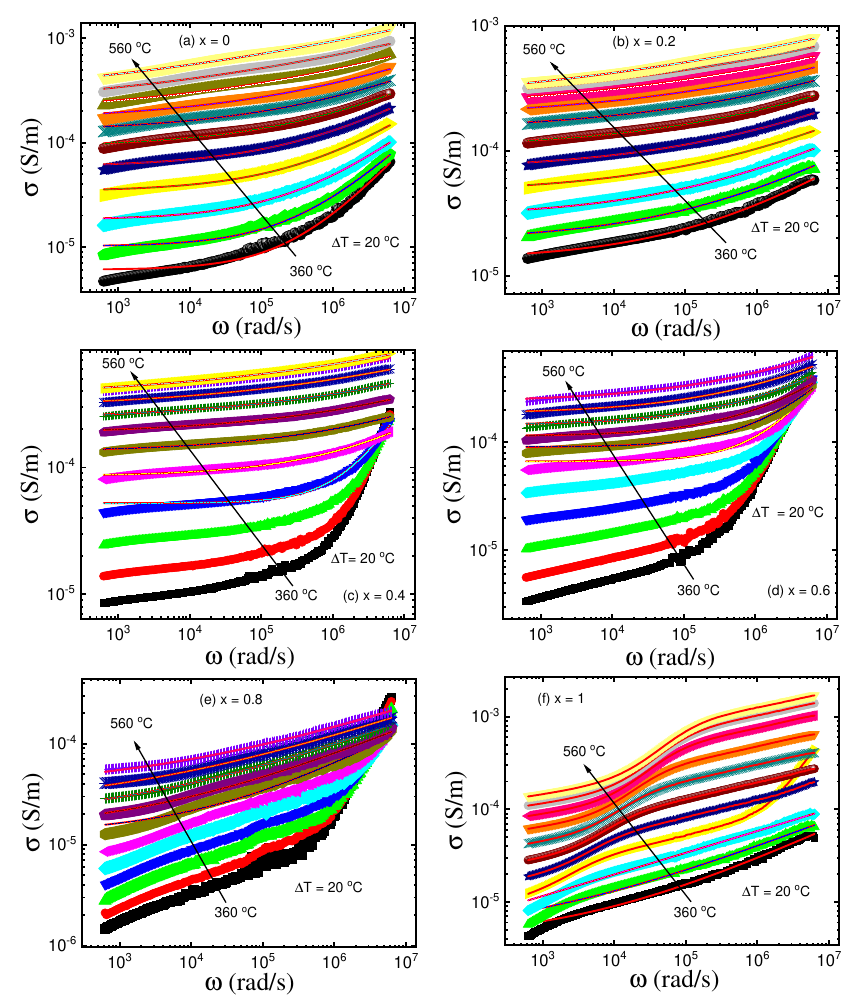}
\caption {(a--f) The angular frequency dependent {\it ac} conductivity for the LaV$_{1-x}$Nb$_x$O$_4$ ($x$ = 0--1) samples at various temperatures between 360\degree C and 560\degree C.}
\label{fit_7} 
\end{figure}

Further, the imaginary part of the impedance ($Z''$) for all the samples is investigated to gain more understanding into various contributing factors, including the effects of electrodes, individual grains, and grain boundaries. Figs.~\ref{fit_4}(a--f) present the imaginary part of the impedance ($Z''$) for all the samples, which is a measure of the phase difference between the applied voltage and the resulting current. In the context of this study, it provides information about the relaxation processes within the material. In the $Z''$ spectra, relaxation maxima are observed, which are characteristic peaks at specific frequencies. These peaks offer details about the contribution of various factors: grain (bulk), grain boundary and ceramic-electrode interface. The samples with $x$ values ranging from 0 to 0.6 exhibit well-defined relaxation peaks in the higher frequency range [see Figs.~\ref{fit_4}(a--d)]. These peaks are attributed to the grain or bulk effect within the samples. In other words, they represent how the individual crystalline grains in the material respond to the applied electrical field. In the samples with $x$ = 0--0.6, a weaker relaxation is also observed in the low-frequency range at higher temperatures. This lower-frequency relaxation is associated with the grain boundary mechanism, which involves how charge carriers move and interact at the interfaces between different crystalline grains. Additionally, the electrode interfaces between the material and the measuring electrodes can contribute to relaxation peaks in the low-frequency region \cite{Bivskup_PRB_05, Lunkenheimer_PRB_04, Lunkenheimer_PRB_02}. As the temperature increases, the relaxation maxima in $Z''$ tend to migrate toward higher frequencies, and their intensity weakens. The temperature-dependent behavior suggests that the relaxation processes within the material are influenced by temperature, with relaxation occurring at higher frequencies as the temperature increases. For the samples with higher Nb substitution ($x$ = 0.8 and 1), stronger relaxation peaks are observed in the lower-frequency regime compared to the samples with $x$ values of 0 to 0.6 as well as at higher frequencies. This suggests that the substitution of Nb$^{5+}$ at the V$^{5+}$ site triggers grain boundary relaxation mechanisms in these samples, especially when $x$ is equal to and greater than 0.8. 

It is important to plot the Nyquist modulus to comprehend the process underlying the relaxation behavior in these samples \cite{TriyonoRSC20}. The modulus spectra [imaginary ($M''$) versus real ($M'$)] that shows the behavior of different semicircular arcs, which are illustrated to distinguish each relaxation mechanism, as shown in Figs.~\ref{fit_6}(a--f) for the samples. One semicircular arc can be seen in the real ($M'$) versus imaginary ($M''$) modulus spectra for all the samples, which indicates that one of the relaxation processes has been severely inhibited by the surface polarization effect. When a small arc is formed at the end of the $M'$ axis, it indicates a minor grain contribution, and when the temperature increases the arc eventually vanishes. This clearly demonstrate how the grain contribution finally diminishes with increase in the temperature. The $M'$ versus $M''$ modulus spectra show that the electrode polarization at high temperature, extracted by the impedance spectroscopy, is drastically inhibited in this condition \cite{31,32}. At the higher frequencies, the grain boundary and electrode ceramic effects can also be seen, which increases with Nb$^{5+}$ concentration and are not visible in impedance plots. In the samples $x$ = 0.4, 0.6 and 0.8, electrode ceramic interface effects are more prominent which may be due to occurrence of mixed phases because of the significant contribution of both Nb$^{5+}$ and V$^{5+}$ elements. 

\begin{figure}[h] 
\includegraphics[width=3.5in]{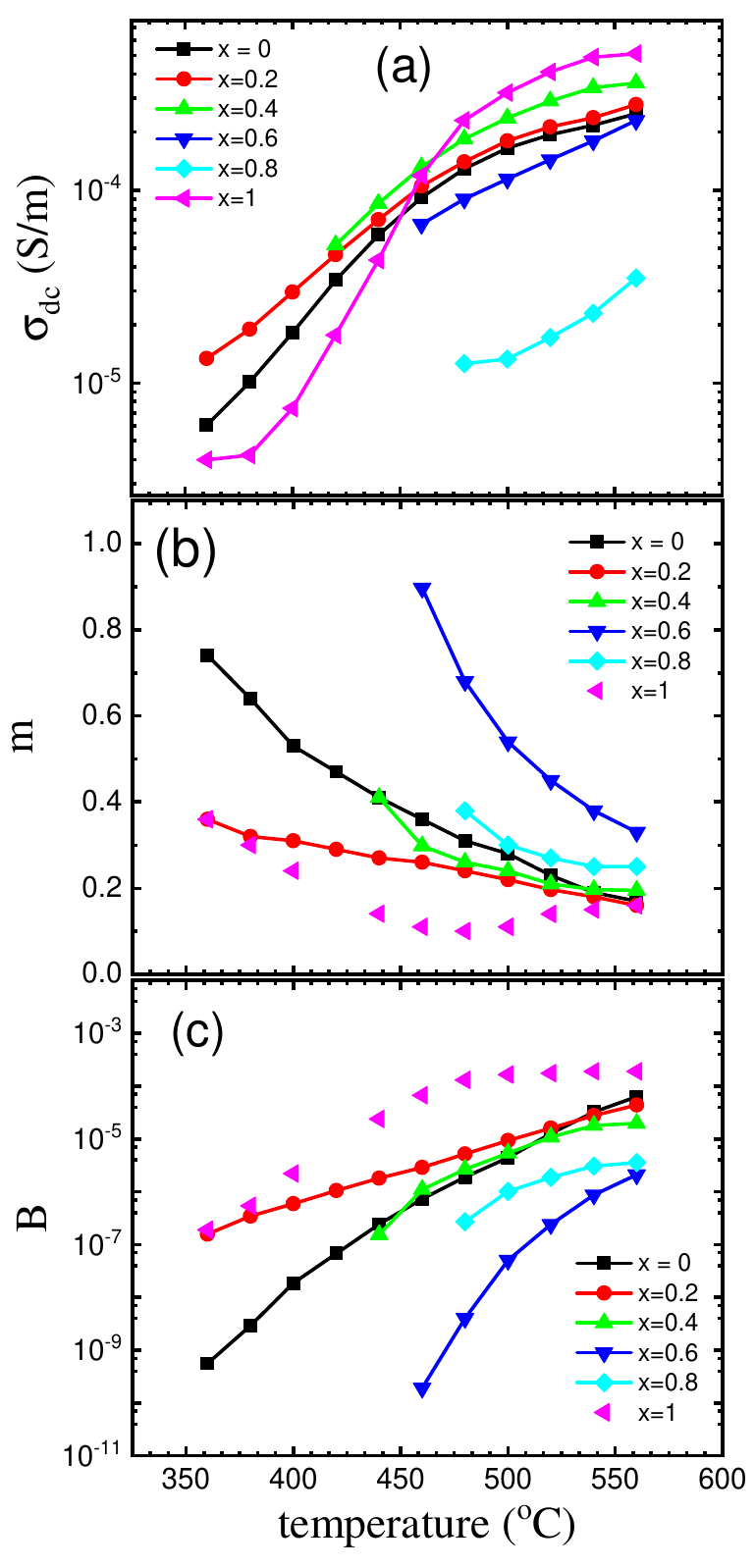}
\caption {The best fit parameters of the $ac$ conductivity using Jonscher power law, (a) the {\it dc} conductivity (b) the exponent $m$, and (c) the constant $B$ as a function of temperature for the LaV$_{1-x}$Nb$_x$O$_4$ ($x$ = 0--1) samples.} 
\label{fit_8} 
\end{figure}

In order to understand this mechanism, we use the relationship 2$\pi \tau f_{max}$=1, where $f_{max}$ represent the frequency at the peak in $Z''$ to determine the dielectric relaxation time ($\tau$). In Fig.~\ref{fit_5}, we plot the $\tau$ as a function of inverse of temperature for all the samples. We find that the relaxation time decreases when the temperature increases and it follow the Arrhenius relation for all the samples: 
\begin{equation}
\tau=\tau_0\exp^{-E_a/k_BT}
\end{equation}
where $E_a$ is the activation energy of the charge carriers, $k_B$ is the Boltzmann’s constant, and $\tau_0$ represents the pre-exponential factor in the relaxation time. The activation energy is typically associated with breaking and forming the chemical bonds, i.e., the minimum amount of energy required for a given relaxation process to activate. The estimated values of activation energy for the $x$ = 0--0.8 samples are found to be 0.98 eV, 0.84 eV, 1.0 eV, 0.88 eV, and 1.41 eV, respectively. The $x$ = 0.8 sample shows maximum activation energy among all, which may be associated with the highest Nb concentration for the coexisting monoclinic-monazite ($P2_1/n$) and tetragonal scheelite ($I4_1/a$) phases, and hence largest average particle size of this samples, observed from the scanning electron microscopy (SEM) \cite{Ashok_Jalcom_23}.  In case of the $x =$ 1 sample, a change in the activation energy is observed near 460\degree C from 1.02 eV to 0.48 eV, which can also be reflected in Fig.~\ref{fit_2}(f), where an additional relaxation peak is observed in this sample. The transformation of the $x =$ 1 sample from the monoclinic fergusonite (space group $I2/a$) structure at the low temperature to the tetragonal scheelite (space group $I4_1/a$) phase at the higher temperature, and hence availability of the large thermal energy of the charge carriers results in this observed reduction in the activation energy above a threshold   temperature of around 460\degree C \cite{TakeiJCG77}. 

In Figs.~\ref{fit_7}(a--f), we present the total conductivity ($\sigma$) of the samples as a function of angular frequency ($\omega$) in the temperature range of 360\degree C--560\degree C. The solid lines represent the best fit of the experimental data using Jonscher Power Law (JPL) equation: 
\begin{equation}
\sigma=\sigma_{dc}+B\omega^m
\end{equation}
where $\sigma_{dc}$ represents the {\it dc} conductivity, $B\omega^m$ represents the {\it ac} conductivity ($\sigma_{ac}$), where both the $B$ and $m$ are the measure of strength of polarization and degrees of interaction of charge carriers, respectively. Both of these quantities are suppose to be dependent on temperature. We find that the $\sigma$ values are strongly frequency dependent and gradually increases at the higher frequencies. Further, the conductivity increases monotonically with increase in temperature for all the samples. Note that at lower temperatures the conductivity increases towards higher frequency; however, it shift towards the saturation at higher temperatures, which indicates a threshold in the conductance mechanism in the materials. At the lower temperature and high frequency region, the conductivity becomes frequency dependent and this frequency dependence increases from $x$ = 0 to 0.8, whereas for the $x$ = 1 sample a weak relaxation in the conductivity is observed in the low frequency region, which is more clearly visible above 400\degree C, which may be attributed to the presence of single monoclinic ($I2/a$) phase in this sample \cite{Ashok_Jalcom_23}. The total conductivity for the $x$ = 0, 0.2 and 1 follows Jonscher Power Law equation in the temperature range of 360\degree C--560\degree C, whereas for the $x$ = 0.4, 0.6 and 0.8 sample it follows the Jonscher Power Law equation above 420\degree C, 460\degree C and 480\degree C temperatures, respectively.   

Finally, in Fig.~\ref{fit_8}(a) we show the variation of the $\sigma_{dc}$ with the temperature where we observe that the $\sigma_{dc}$ increases non-linearly in the temperature range of 360\degree C--560\degree C, which shows the semiconducting/insulating nature of the materials. There have been various models reported to understand the conductivity of different complex oxide samples on the basis of exponent $m$ as well as the polaron tunneling. It is important to note that a local lattice deformation can result due to the addition of charge transporters to a site in a solid. Further, a polaron can be created by the arrangement of electrons as well as the local distortion. In the current study, we find that the $m$ values decreases with temperature for all the samples, except for $x$ = 1, which shows the decreasing valley near 480\degree C and the subsequent increase in value indicates the presence of the overlapping large polaron tunneling (OLPT) conduction mechanism \cite{33,34}. In this mechanism, the polaron clouds extend considerably beyond the typical interatomic distances \cite{59}. As the temperature increases, the $m$ values shown in Fig.~\ref{fit_8}(b) exhibits a linear ascent until it reaches approximately 540\degree C, elucidating the behavior of non-overlapping small polaron tunneling (NSPT) \cite{35}. During this phase, the polarons are confined to the extent that their distortion clouds do not intersect. Consequently, the electron's tunnel between states near the Fermi level, and the tunneling energy remains unaffected by inter-site distances \cite{59}. The $m$ value eventually levels off at around 560\degree C, a phenomenon explicable by quantum mechanical tunneling (QMT) \cite{36}. Notably, the temperature independence of $m$ in this model distinguishes it \cite{60}. This characteristic underscores that conduction occurs primarily through the tunneling of electrons between defect states near the Fermi level, without the generation of polarons. Similarly, the values of $B$ shown in Fig.~\ref{fit_8}(c) increases non-linearly with temperature for all the samples, except for the $x$ = 1, which shows the saturation near 560\degree C, which means the polarization strength is also getting saturated at higher temperatures, which is found to be consistent with the dielectric behavior, as shown in Fig.~\ref{fit_2}. 

\section{\noindent ~Summary and Conclusions}

We measured the dielectric constant and tangent loss at temperatures ranging from 25\degree C to 600\degree C and a wide range of frequencies (100 Hz to 1 MHz) for LaV$_{1-x}$Nb$_x$O$_4$ ($x$ = 0--1) samples. We found that the dielectric constant increased with temperatures, but decreased in magnitude with frequency having the value $\approx$10$^4$ at 100~Hz and 600\degree C. The iso-electronic substitution of Nb$^{5+}$ at V$^{5+}$ site did not significantly affect the magnitude of the dielectric constant. The anomalous behavior in $\epsilon_r'$(T) curves have been observed for the $x$ = 0.2--0.8 samples due to their mixed structural phases, whereas no relaxation peak was observed for the $x$ = 0 sample, and a weak relaxation was seen for the $x$ = 1 sample at the higher temperatures. The mixed phase in the $x$ = 0 to 0.8 samples show the significantly higher values of the tangent loss as compared to the single phase $x =$ 0 and 1 samples. The real part of the impedance ($Z'$) spectroscopy decreased with temperature and frequency, and a weak relaxation was observed in all the samples with negligible changes in magnitude with $x$. The imaginary part of the impedance ($Z''$) exhibited strong relaxation peaks that shifted toward higher temperatures with frequency, which described the grain, grain boundary, and electrode effects in the samples. The $Z''$--$f$ plots exhibited similar behavior for the $x$ = 0 to 0.6 samples, but for the $x$ = 0.8 and 1, the relaxation was observed in the lower frequency region, indicating the grain boundary effect along with the grain effect. The activation energy of the relaxation processes for the $x$ = 0--0.6 samples found to be in the range of 0.8 to 1.0~eV, whereas for the $x$ = 0.8, the activation energy was calculated to be around 1.4~eV due to highest Nb concentration in the multiphase samples. Further, for the $x$ = 1 sample, two values of the activation energy were observed, i.e., around 1.0~eV and 0.5~eV, which indicated the transition near 460\degree C. We measured the total conductivity (10$^{-3}$ to 10$^{-5}$~S/m) with varying frequency in the temperature range of 360\degree--560\degree C and fitted it with the Jonsher power law. The fitted {\it dc} conductivity behavior showed the semiconducting characteristics of all the samples, with $x$ = 0.8 having the lowest conductivity. The analysis showed that all the samples exhibited OLPT mechanism, whereas the $x$ = 1 exhibited a transition near 480\degree C, which was also reflected in the tangent loss plots. As an exception, we find NSPT and QMT mechanisms for the $x$ = 1 sample at higher temperatures. 







\section*{\noindent ~Acknowledgments}

V.N.T. and Ajay Kumar thank DST (project: DST/TMD/MECSP/2K17/07) and UGC, respectively, for the fellowship. RSD thanks the IIT Delhi for supporting the development of temperature dependent dielectric set-up in the group. RSD acknowledges SERB--DST for the financial support through a core research grant (project reference no. CRG/2020/003436).

\end{document}